\newcommand{\ben}{\begin{equation}}
\newcommand{\een}{\end{equation}}
\newcommand{\bea}{\begin{eqnarray}}
\newcommand{\eea}{\end{eqnarray}}
\def\bra#1{\langle#1\vert}
\def\ket#1{\vert#1\rangle}

\def\sss{\scriptscriptstyle\rm}

\def\1s{_{1,\sss S}}
\def\2s{_{2,\sss S}}
\def\x{_{\sss X}}

\def\s{_{\sss S}}
\def\xc{_{\sss XC}}

\def\Hxc{_{\sss HXC}}

\def\H{_{\sss H}}

\def\ext{_{\rm ext}}

\def\br{{\bf r}}

\def\eps{{\epsilon}}

\documentclass{ar-1col}
\usepackage[numbers]{natbib}
\usepackage{url}
\jname{Xxxx. Xxx. Xxx. Xxx.}
\jvol{AA}
\jyear{YYYY}
\doi{10.1146/((please add article doi))}

\begin{document}

\markboth{Maitra}{Double and Charge-Transfer Excitations in TDDFT}

\title{Double and Charge-Transfer Excitations in Time-Dependent Density Functional Theory}

\author{Neepa T. Maitra $^1$
\affil{$^1$Department of Physics, Rutgers University at Newark, Newark, NJ 08904, USA; email: neepa.maitra@rutgers.edu}}

\begin{abstract}
Time-dependent density functional theory has emerged as a method of choice for calculations of spectra and response properties in physics, chemistry, and biology, with its system-size scaling enabling computations on systems much larger than possible otherwise. 
While increasingly complex and interesting systems have been successfully tackled with relatively  simple functional approximations, there has also been increasing awareness that these functionals tend to fail for certain classes of approximations.  
I review the fundamental challenges the approximate functionals have in describing double-excitations and charge-transfer excitations, which are two of the most common impediments for the theory to be applied in a black box way. At the same time, I describe the progress made in recent decades in developing functional approximations that give useful predictions for these excitations. 
\end{abstract}

\maketitle

\tableofcontents

\section{INTRODUCTION}
The response properties of an atom, molecule, or solid are a crucial aspect of its characterization. In ``poking" a system with a perturbation, be it sunlight, or a weak laser, or a collision with a small particle, one causes a gentle disturbance in its constituents whose consequent dynamics reveals a wealth of information about interactions in the system. Depending on how the time-scales involved in the perturbation compare to the effective reaction times of the constituents, different types of correlation can be revealed: sometimes only the electronic system is probed, while other times it is the vibrational or rotational motion of the ions that responds, or a combination. When the perturbation is weak enough that the system response scales linearly with the strength of the perturbation, the response at different frequencies and different wavevectors, i.e. the absorption spectrum, shows peaks at excitation energies of the system, whose strength indicates the transition probability from the ground to that excited state.

Thus, the spectrum, meaning the excitation energies, and their oscillator strengths, provides a unique signature of the system. Theoretical computations attempt to forge these signatures, in order to make predictions of dynamics and processes, to identify or characterize an experimental spectrum, and to design new materials with some desired spectral property. Solving Schr\"odinger's equation for the many-electron system however scales exponentially with the number of electrons, so some kind of approximation is ultimately called for.  
Time-dependent density functional theory (TDDFT) provides a route to computing the absorption spectra which has achieved overall the best balance between accuracy and efficiency. Formulated in 1984~\cite{RG84}, with the linear response framework developed in 1995~\cite{C95,PGG96}, the past 25 years have seen some very exciting applications for response properties. The functional approximations that were used in its early days are often still used today. However, there are certain excitations for which these approximations perform poorly, and more sophisticated approximations are needed. Perhaps the most important cases of this are double-excitations and charge-transfer excitations.

Both double- and charge-transfer excitations have not merely a theoretical interest, but also have important practical importance in a number of situations, from photovoltaic design to organic molecules to biochemical processes. This is particularly true when there is coupling to ionic motion and the molecule explores a wide range of geometries. There has therefore been tremendous effort to describe these excitations, not just within TDDFT, but also with other many-body wavefunction methods. I note that these excitations can be challenging also for wavefunction methods: for example, the error in equation-of-motion coupled-cluster methods is related to the amount of double-excitation in the transition~\cite{LBSCJ19}. The general problem of computing accurate excited states has inspired the development of the QUESTDB database, which has more than 500 highly accurate vertical excitations of different natures~\cite{VSCL21} in small and medium-sized molecules, using only computational data. 

This review focusses on the TDDFT description of these two classes of excitations. I begin with a review of how linear response calculations proceed in TDDFT before first addressing the challenge, and some solutions, in obtaining double excitations in Sec.~\ref{sec:doubles}, and charge-transfer excitations in Sec.~\ref{sec:CT}. I provide an outlook in Sec.~\ref{sec:outlook}.


\section{TDDFT Linear Response}
\label{sec:TDLR}
In a nutshell, TDDFT is an exact reformulation of the quantum dynamics of many-body systems, where a non-interacting system of $N$ electrons reproduces the one-body density of the true system~\cite{RG84,Carstenbook,TDDFTbook12,M16,CH12},  $n(\br, t) = N\sum_{\sigma, \sigma_2 ... \sigma_N} \int \vert \Psi(\br\sigma, \br_2\sigma_2 ...\br_N\sigma_N)\vert^2 \vert d^3 r_2...d^3 r_N$. Instead of finding the true correlated wavefunction, a problem that scales exponentially with the number of electrons, one needs only find a set of $N$ orbitals that satisfy the time-dependent Kohn-Sham (KS)  \begin{marginnote}[]\entry{KS}{Kohn-Sham}\end{marginnote} equations, 
\ben
\left(-\frac{\hbar^2}{2m} \nabla^2 + v\s(\br,t)\right)\phi_k(\br, t) = i \partial_t \phi_k(\br,t)
\label{eq:tdks}
\een
where the KS potential $v\s(\br, t) = v\ext(\br,t) + v\H[n](\br, t) + v\xc[n; \Psi_0, \Phi_0](\br,t)$ with the Hartree potential $v\H[n](\br, t) = \int \frac{n(\br', t)}{\vert \br - \br'\vert}d^3r'$ being the classical electrostatic repulsion of the electron distribution, and $v\xc[n;  \Psi_0, \Phi_0](\br,t)$ is the exchange-correlation (xc) \begin{marginnote}[]\entry{xc}{exchange-correlation}\end{marginnote} potential which is defined such that the time-evolving density of the occupied orbitals is identical with the true density,  $n(\br,t) = \sum_{k, occ.}\vert \phi_k(\br,t)\vert^2$. The xc potential is a functional of the density, including its history, as well as the initial interacting state $\Psi_0$ and non-interacting state $\Phi_0$; when the dynamics begins in the ground-state, as in the linear response regime, the initial-state dependence is usurped into the density-dependence by the Hohenberg-Kohn theorem of ground-state DFT~\cite{HK64},  and the xc potential may be simply written as $v\xc[n](\br,t)$.

 In practise, of course, $v\xc$ is unknown and approximations are made, most of which simply insert the instantaneous density into a ground-state approximation. This ``adiabatic approximation" \begin{marginnote}[]\entry{adiabatic approximation}{no memory, i.e. instantaneous dependence on the density as if in a ground-state}\end{marginnote}thus
completely neglects both the history- and initial-state-dependence, yet has led to many useful predictions in both the linear response as well as non-perturbative regime (see Refs. e.g. in~\cite{M16}). In the past couple of decades there has been increased understanding of where and why the functional approximations fail, especially in the linear response regime, such that users know when to trust their TDDFT results using standard functionals, and when to be cautious. More sophisticated functionals have been developed from first-principles, which, while computationally more involved, deliver more reliable results for classes of excitations for which the standard approximations fail. Two of these classes, double- and long-range charge-transfer excitations, are the focus of this review. 

When applied to linear response, the formalism and its functionals simplify. 
First-order perturbation theory  gives the linear response of the density to a perturbation $\delta v(\br, t)$:
\ben
n^{(1)}(\br,t) = \int_0^\infty dt' \int d^3 r' \chi(\br t, \br' t') \delta v(\br't')
\een
where in the frequency-domain the density-density response function $\chi$  \begin{marginnote}[]\entry{$\chi$}{density-density response function}\end{marginnote}has poles at the exact frequencies $\Omega_j = E_j - E_0$ and the residues give transition densities between the true ground $\Psi_0$ and excited $\Psi_j$ many-body states:
\ben
\chi(\br,\br',\omega) = \sum_j \frac{\langle \Psi_0\vert\hat{n}(\br)\vert\Psi_j\rangle\langle \Psi_j\vert\hat{n}(\br')\vert\Psi_0\rangle}{\omega - \Omega_j + i 0^+}  + c.c. (-\omega)
\label{eq:chi}
\een
where $\hat{n}(\br) = \sum_i \delta(\br - \br_i)$ is the one-body density operator. 
The notation $c.c.(-\omega)$ denotes the complex conjugate of the first term at frequency $-\omega$. 
Using the fact that the time-dependent KS system yields the same density as the interacting system,  TDDFT provides an expression for the response function $\chi$ that bypasses finding the excited states~\cite{PGG96,C96}:
\ben
\chi[n](\br,\br',\omega) = \chi\s[n](\br,\br',\omega) + \int d^3r_1 d^3r_2 \chi\s(\br, \br_1,\omega)\left(\frac{1}{\vert \br_1 - \br_2\vert} + f\xc(\br_1,\br_2,\omega)\right) \chi(\br_2, \br',\omega)
\label{eq:dyson}
\een
where 
\ben
\chi\s[n](\br,\br',\omega) = \lim_{\eta \to 0^+}\sum_{k,j}(f_k - f_j) \delta_{\sigma_k,\sigma_j}\frac{\phi_k^*(\br)\phi_j(\br)\phi_j^*(\br')\phi_k^(\br')}{\omega - (\epsilon_j - \epsilon_k)+ i\eta}
\label{eq:chis}
\een
is the time-frequency Fourier transform of $\chi\s(\br,\br',t-t') = \frac{\delta n(\br,t)}{\delta v\s(\br',t')}$,
and the xc kernel\begin{marginnote}[]\entry{$f\xc[n](\br,\br',\omega)$}{xc kernel}\end{marginnote} is that of $f\xc[n](\br,\br',t-t') = \frac{\delta v\xc[n](\br,t)}{\delta n(\br' t')}$. In Eq.~\ref{eq:chis}, $f_i$ are Fermi occupation numbers of the orbital $i$ in the ground-state KS  determinant. 
 Eq.~\ref{eq:chi} involves a sum over excited states; the $j = 0$ term in the first and second term cancel. 

Eq.~\ref{eq:dyson} is the central equation in TDDFT linear response~\cite{PGG96}, showing how the excitation energies and transition densities of the true interacting system are related to those of the non-interacting KS system  and the xc kernel. 
In practise, although this  is used directly for extended systems, for molecules a matrix formulation is used~\cite{C95,C96}. There are several versions which are all essentially equivalent~\cite{Carstenbook,GM12chap,GPG00}, and involve solving for eigenvalues and eigenvectors of a matrix in the basis of KS single excitations. 
In the Casida formulation~\cite{C95,C96,GPG00} we have
\ben
R(\Omega_j)F_j = \Omega_j^2F_j
\een
(provided the orbitals are chosen real), where
\ben
R_{qq'}(\omega) = \omega_q^2 \delta_{qq'} + 4 \sqrt{\omega_q \omega_{q'}}\int d^3r d^3r' \Phi_q(\br) \left(\frac{1}{\vert \br -\br'\vert} + f\xc(\br, \br', \omega)\right)\Phi_{q'}(\br')
\label{eq:matrix}
\een
with $q = i\to a$ denoting a single KS excitation from an occupied orbital $i$ to unoccupied $a$, and $\Phi_q(\br) = \phi_i^*(\br)\phi_a(\br)$ a KS transition density. 
In Eq.~\ref{eq:matrix} we have restricted ourselves to spin-saturated closed-shell systems for simplicity; the spin-dependent version of these equations yield singlet-triplet splittings. The eigenvalues of $R$ yield the excitation energies $\Omega_j$ while the true transition densities and oscillators are related to the eigenvectors $F_j$. 

Two truncations of Eq.~\ref{eq:dyson} or Eq.~\ref{eq:matrix} are particularly useful tools for analysis. In the ``small matrix approximation" (SMA) \begin{marginnote}[]\entry{SMA}{small matrix approximation}\end{marginnote} we focus in on one KS excitation and assume it has negligible coupling to the other excitations. Then~\cite{PGG96,GPG00,AGB03,GM12chap},
\ben
\Omega^{\rm SMA} = \sqrt{\omega_q^2 + 4\omega_q \int d^3r d^3r' \Phi_q(\br) f\Hxc(\br,\br',\omega)\Phi_q(\br')}\,.
\label{eq:SMA}
\een
where the notation Hxc denotes Hartree-xc, $f\Hxc = \frac{1}{\vert \br - \br'\vert} + f\xc$. 
If, additionally, the correction to the KS excitation frequency $\omega_q$ is itself much smaller than $\omega_q$, taking a Taylor expansion yields the ``single pole approximation" (SPA) \begin{marginnote}[]\entry{SPA}{single pole approximation}\end{marginnote}
\ben
\Omega^{\rm SPA} = \omega_q + 2\int d^3r d^3r' \Phi_q(\br) f\Hxc(\br,\br',\omega)\Phi_q(\br')\,.
\label{eq:SPA}
\een

In principle, use of the exact ground-state xc potential and exact xc kernel in Eq.~\ref{eq:dyson} or Eq.~\ref{eq:matrix}  would yield exact excitation energies and transition densities of the physical system. However both these ingredients are unknown and need to be approximated in practise. The choices for the ground-state functional are enormous~\cite{B12,Becke14,Truhlar16,Goerigk19}, vary hugely in their degree of empiricism as well as in their computational cost. For the xc kernel, the adiabatic approximation for the xc potential translates to a frequency-independent $f\xc$ which follows from: $v^A\xc [n](\br,t) = v^{\rm g.s.}\xc[n(t)](\br)$, then $f\xc[n](\br,\br', t -t') = \frac{\delta^2 E\xc[n]}{\delta n(\br)\delta n(\br')} \delta(t - t')$, so 
\ben
f\xc^A[n](\br,\br',\omega) = \left.\frac{\delta^2 E\xc[n]}{\delta n(\br)\delta n(\br')}\right\vert_{n=n(\br)}
\een

In the following sections, we will analyze what the challenge is with linear response TDDFT for double-excitations and charge-transfer excitations, as well as possible solutions, from the perspectives of both Eq.~\ref{eq:dyson} and Eq.~\ref{eq:matrix}. Both expressions demonstrate the key role played by the xc kernel in producing the response of the interacting system from that of the KS one, but also the key role played by the bare KS excitations themselves as a zeroth order starting point for the TDDFT machinery. 
We will see that while the structure of the kernel is the crucial aspect in capturing double-excitations and charge-transfer excitations between open-shell fragments, a good approximation for the ground-state kernel is a crucial aspect of getting charge-transfer between closed-shell fragments correct. 

Before doing so, we note that two other formulations of linear response within TDDFT, which may be more computationally efficient for certain situations. The Sternheimer approach, also known as density perturbation theory, or coupled perturbed KS, avoids the calculation of unoccupied states by instead considering perturbations of the occupied KS orbitals in frequency-domain~\cite{Sternheimer54,ABMR07}. Also avoiding unoccupied orbitals, one can Fourier-transform the real-time propagation of occupied orbitals under a weak perturbation, often using a $\delta$-kick in time to uniformly stimulate the entire spectrum~\cite{YNIB06}.


\section{Double excitations}
\label{sec:doubles}
\begin{figure}[h]
\includegraphics[width=\textwidth]{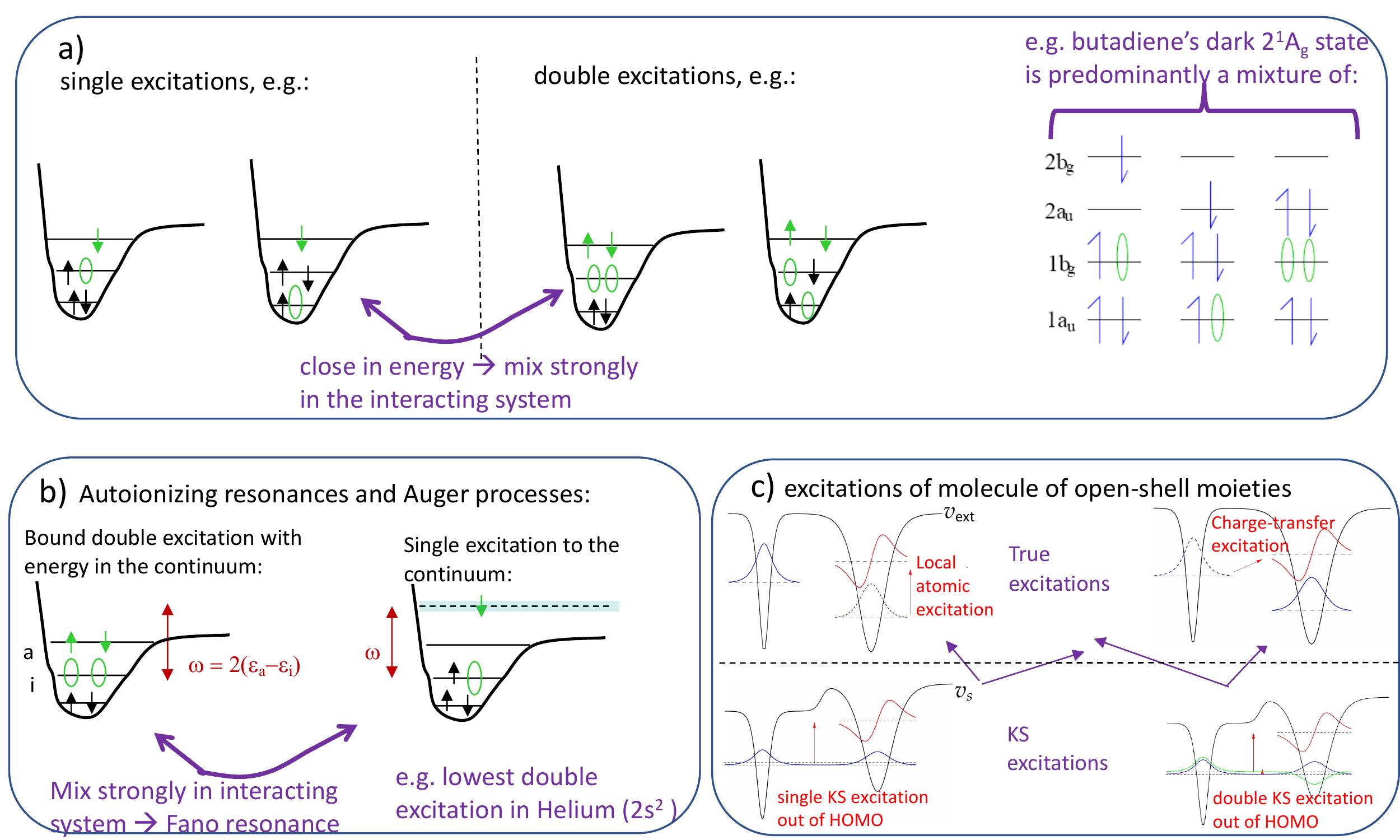}
\caption{Double-excitations in different contexts: a) Single vs double excitations, defined with respect to a single-determinant reference such as KS, and the example of the butadiene $2^1A_g$ state, which is a mixture of two single and one double excitation out of of the KS single Slater determinant reference. b) Auto-ionizing resonances often involve double excitations, where, in a single-particle reference, a double-excitation to a bound orbital has an energy that lies in the continuum. Once electron-interaction is accounted for, the state turns into a resonance, and TDDFT with a frequency-dependent kernel can give approximate linewidths~\cite{EGCM11}. c) Excitations of a stretched molecule exemplify the ubiquity of double-excitations when static correlation is present.}
\label{fig:doublesketch}
\end{figure}

First, what {\it is} a double excitation? The term is a short-hand for a state of double-excitation character, and has meaning in the context of non-interacting reference systems such as Hartree-Fock or KS DFT~\cite{EGCM11}. In these systems one has $N$ orbitals that are occupied in the ground state, where $N$ is the number of electrons, and an infinite number of virtual orbitals.  A doubly-excited Slater determinant is when  two electrons are promoted out of  occupied orbitals into two virtual orbitals, and a double-excitation of the true interacting state is then one which has a significant proportion of doubly-excited determinants in an expansion of the true correlated state using the non-interacting reference states (see also (\textbf{Figure \ref{fig:doublesketch}}). Clearly, the details of the expansion coefficients  and the orbitals themselves are dependent on which non-interacting reference is chosen. Whether a given state should be classified as a single excitation (usually meaning a linear combination of single-excitations, again with respect to a chosen non-interacting reference) or a double- or multiple-excitation, generally then depends on the choice of the reference. Note though, that for a given reference, a well-defined and unambiguous classification can be made by following the states as the interaction is slowly turned down to zero. For DFT references, this is done along the adiabatic connection curve, and  the procedure is described in Ref.~\cite{ZB04}
 (see also discussion~\cite{FD2020}).

But this definition relying on the notion of a non-interacting reference means that whether a state is classified as being a double excitation or not can lose intrinsic meaning. A double excitation using a single  determinant reference such as Hartree-Fock or KS DFT, may appear as single excitations from a multi-reference ground-state, or, if excited-state orbital-relaxation is accounted for, such has been discussed for the case of butadiene for example~\cite{LBSCJ19,ST17,BGG18}. However, within TDDFT no such ambiguity arises, because the KS ground-state is a Slater determinant (except in cases of strict degeneracy), and the excitations are obtained within a fixed basis of occupied and unoccupied orbitals once the ground-state is determined. i.e. there is no orbital-relaxation as such. Thus, the double-excitation character of the state is well-defined in TDDFT, determined by the path of Ref.~\cite{ZB04}.

Now we turn to how double excitations appear in the linear response spectrum. In fact, they are completely absent in the KS linear response, and it 
 is purely through their interaction with a single excitation that they appear in the true response function, as we will now explicitly demonstrate. We expand the true interacting states $\ket{\Psi_j}$ in the complete set formed by KS determinants: 
 \ben
 \ket{\Psi_j} = C^j_0\ket{\Phi_0}  + \sum_{q}C^j_{q}\ket{\Phi_q} + \sum_D C^j_{D}\ket{\Phi_D} +...
 \label{eq:psi-exp}
 \een
 where $q = i\to a$ represents all single excitations out of the KS ground-state determinant $\Phi_0$, and $D = ( i \to a, j \to b)$ represents all double excitations, etc. 
 The numerator of Eq.~\ref{eq:chi} involves matrix elements between the ground and excited states of the density-operator $\hat{n}(\br)$, which, being a one-body operator, has only non-zero elements between determinants which differ by at most one orbital. For this reason, even before considering $\chi$, it is instructive to apply Eq.~\ref{eq:chi}  to the non-interacting KS system, where the numerator involves $\bra{\Phi_j}\hat{n}(\br)\ket{\Phi_0}$: only the single excitations $\bra{\Phi_j} = \bra{\Phi_q}$ give a non-zero contribution. Physically, this is to be expected since a double-excitation in a non-interacting system would mean that two electrons are excited, a process that would require two photons, and so scaling quadratically in the perturbation strength, not linearly. 
 
 Returning to the interacting system, it follows from above that if the true ground-state was weakly correlated and well-approximated by a Slater-determinant, then excited states of double-excitation character $\ket{\Psi_j}$ contribute to the linear response solely through their single-excitation component $\ket{\Phi_q}$ in the expansion of Eq.~\ref{eq:psi-exp}. More generally, putting the expansion Eq.~\ref{eq:psi-exp}  for $\ket{\Psi_0}$ and $\ket{\Psi_j}$ into Eq.~\ref{eq:chi}, we see that the $\ket{\Phi_0}$ component of the interacting ground-state $\ket{\Psi_0}$ gives a non-zero contribution only through matrix elements with the single-excitation components $\ket{\Phi_q}$ of the excited state $\ket{\Psi_j}$, while the single-excitation component of the ground-state ($\ket{\Phi_q}$ in the expansion of $\ket{\Psi_0}$) gives a non-zero contribution through the double-excitation component $\ket{\Phi_D}$ of the excited-state, while the double-excitation component of the ground-state connects to the single-excitation and any triple-excitation components of the excited-state.  Likewise, the double-excitation component of the excited state connects only to single-excitation  and any triple-excitation components of the ground-state. As triple-excitations are generally much higher in energy, we see that the double-excitations really contribute only through couplings with the single-excitations. 
 There are clearly more poles in the interacting $\chi$ than in the KS $\chi\s$: $\chi$ has poles at true states that are linear combinations of single, double, and higher excitations, while $\chi\s$ has poles only at single excitations.

Given that in TDDFT $\chi$ is obtained from $\chi\s$ through Eq.~\ref{eq:dyson}, the appearance of double-excitations in $\chi$ depends {\it entirely} on the xc kernel. That is, unlike single-excitations, there is no zeroth order approximation to the double-excitation that can be extracted from the response function or in Casida's matrix. One could resort to taking sums of the KS single excitations as a zeroth order approximation, but such a term does not naturally arise in the TDDFT linear response formalism. 
Further, to generate more poles than the KS system has, the xc kernel must be strongly frequency-dependent. 
Another way to see this is through the matrix equation, Eq.~\ref{eq:matrix}. Since this is a matrix spanning single-excitations, the only way that information about a double-excitation can enter is implicitly through $f\xc$. Because the interacting system has a larger number of excitations than the non-interacting ones, the equation must represent a non-linear rather than linear eigenvalue problem, which arises due the frequency-dependence of the xc kernel~\cite{JCS96,TH00,MZCB04,EGCM11,AL20}. 

The lack of double-excitations in the adiabatic approximation was noted soon after TDDFT linear response was formulated~\cite{JCS96,TH00}, where it was also suspected that 
including frequency-dependence would unveil them. 
Ref.~\cite{TK09} demonstrated numerically the need for frequency-dependence to capture double-excitations by showing that the ``adiabatically-exact" approximation misses their peaks in the absorption spectrum. The adiabatically-exact approximation is the best that an adiabatic approximation could hope to be, since it inputs the instantaneous density into the exact ground-state xc functional. This can only be done for model systems where the exact ground-state xc functional is numerically accessible, and Ref.~\cite{TK09} ran real-time propagation on some one-dimensional two-electron systems in the linear response regime to demonstrate this.

An explicit computation of the frequency-dependence of the exact xc kernel is quite a computational feat, given that it is effectively solving an inverse problem which is very sensitive to small errors. Yet it has been achieved~\cite{TK14,EG19,WEG21} on model systems, and results verify the simple pole structure of the xc kernel near double or multiple excitations that was postulated in a simple model in Ref.~\cite{MZCB04} (Sec.~\ref{sec:dtddft}). Ref.~\cite{TK14} performed real-time calculations of a kick-perturbation that is localized in space and time,  to find effectively the functional derivatives in  $\chi(\br,\br',t-t')$ and $\chi\s(\br,\br',t-t')$, then Fourier-transforming to the frequency-domain, to reveal a full spatial and frequency-dependency of the kernel, $f\xc = \chi\s^{-1} -\chi^{-1} - 1/|\br - \br'|$. Refs.~\cite{EG19,WEG21} worked directly in the frequency domain to construct the true and KS response functions. In both approaches, regions of small density need much care; a thorough analysis together with different ways to deal with this can be found in Ref.~\cite{WEG21}. An interesting issue that arose is the ``gauge-freedom" of the xc kernel~\cite{WEG21,HB08,HG12}: adding functions $g(\br,\omega)$ (independent of $\br'$), $h(\br',\omega)$ (independent of $\br$), or a spatially-independent term, to $f\xc(\br,\br',\omega)$ has no effect in the Dyson equation Eq.~\ref{eq:dyson}. This reflects the fact that the physics is invariant under a spatially uniform but possibly time-dependent shift of the potential. 
The works above utilized the soft-Coulomb interaction between electrons that is often used in one-dimensional models. An exact analytic expression for the xc kernel has also been shown on two different models with different interactions: one for a ring geometry with a squared cosine interaction~\cite{RNL13}, and the other for a lattice system, the (a)symmetric Hubbard dimer~\cite{FM14,AG02,TR13,CFMB18}.

\subsection{Dressed TDDFT}
\label{sec:dtddft}
\begin{figure}[h]
\includegraphics[width=4in]{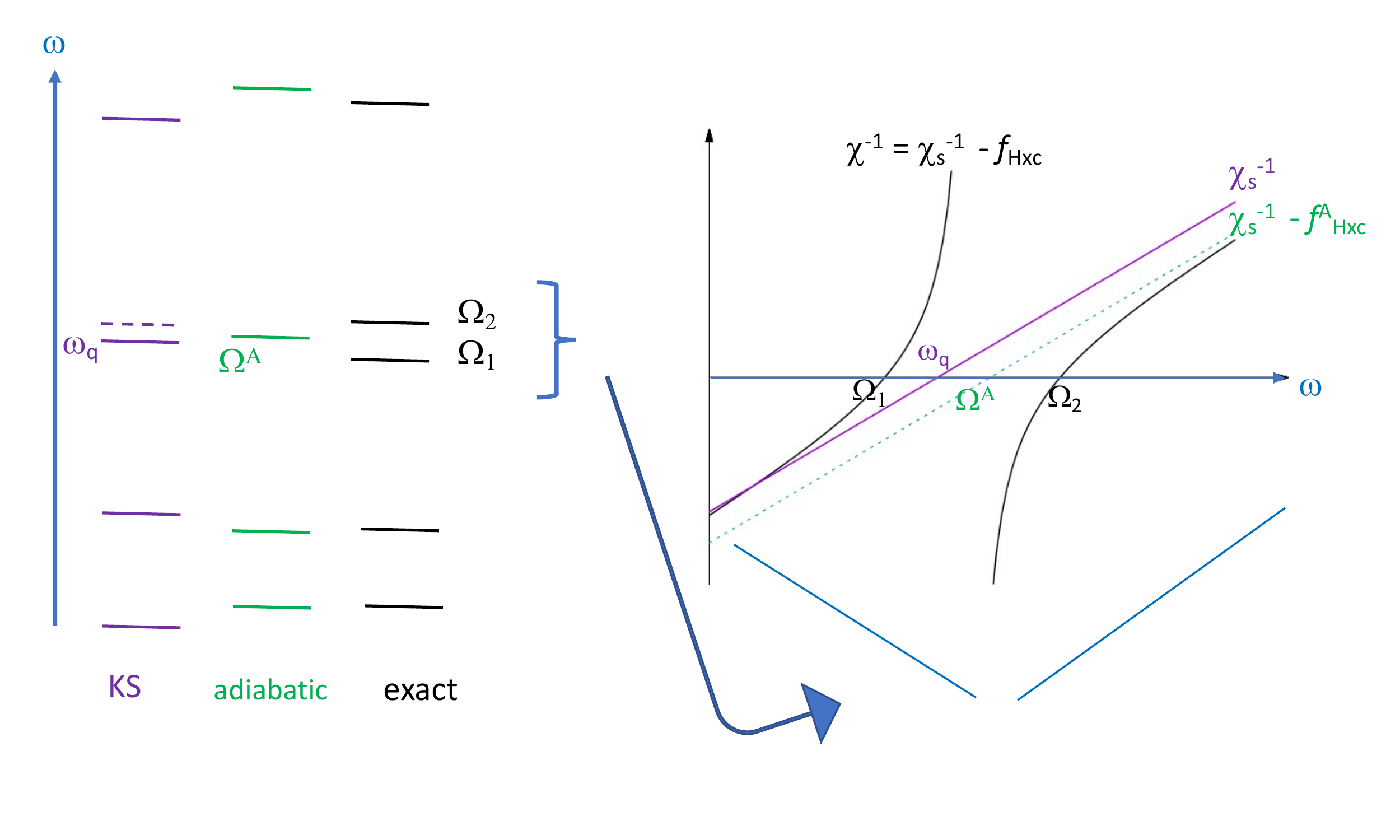}
\caption{Cartoon showing how the frequency-dependence in the dressed kernel generates an extra pole}
\label{fig:kernelsketch}
\end{figure}

To reveal the hidden double-excitations in $f\xc$, Ref.~\cite{MZCB04} considered a simple idealized situation where, in the KS system, one double-excitation is close in energy to a single-excitation and both are far away from all other excitations, such that in the frequency range near these two states, the KS response function has a single pole. 
 Electron interaction mixes these two excitations such that there are two excitations in the interacting system that are linear combinations of this KS single and double, each contributing a distinct pole in the response function $\chi$. Motivated by the expression obtained from solving  Eq.~\ref{eq:matrix} in this subspace for $f\xc$, Ref.~\cite{MZCB04}
asserted the dressed SPA (dSPA) for the xc kernel, 
 \ben
\int d^3r d^3r' \Phi_q(\br) f\xc^{\rm dSPA}(\br,\br',\omega)  \Phi_q(\br') = \int d^3r d^3r' \Phi_q(\br) f\xc^A(\br,\br')  \Phi_q(\br') + \frac{\vert H_{qD}\vert^2/2}{\omega - (H_{DD} - H_{00})}
\label{eq:doubleskernel}
\een
where the second term gives a frequency-dependent correction to a chosen adiabatic approximation (1st term), involving Hamiltonian matrix elements with the KS single (q), double (D), and ground-state (0). (A sketch is shown in \textbf{Figure \ref{fig:kernelsketch}})
The kernel was proposed as an {\it a posteriori} correction to the adiabatic approximation, applied to just the particular KS single excitation that lies near a double-excitation. Its derivation relies on the idea that the interaction of these two excitations with the other KS excitations in the system is far weaker than with each other. If several KS single excitations mix significantly with a double-excitation, then the dressing can be applied in a matrix spanned by those singles, in a dressed Tamm-Dancoff scheme, as was done for small polyenes in Ref.~\cite{CZMB04,MW09,MMWA11} (see also  panel a) in \textbf{Figure \ref{fig:kernelsketch}}). 
Ref.~\cite{GB09} unveiled the spatial-dependence of the kernel, and, using the common energy denominator approximation, could approximately account for the effect of the entire spectrum on the coupled singly and doubly excited states.

Several related and more rigorously-based approaches have led to kernels of essentially the same form as Eq.~\ref{eq:doubleskernel}. Ref.~\cite{C05} used the equation-of-motion superoperator approach to derive a polarization propagator equation, separating out the adiabatic and non-adiabatic contributions. The non-adiabatic part was shown to reduce to Eq.~\ref{eq:doubleskernel} in the special case where the ground-state is closed-shell, while in the general case, it provides an extension of dressed TDDFT to open-shell doublets. Ref.~\cite{RSBS09} built an xc kernel from contracting the four-point Bethe-Salpeter equation of many-body theory to the two-point one of TDDFT~\cite{BSOSR05,GORT07}. Usually a static approximation is used for the many-body kernel but Ref~\cite{RSBS09} showed that a frequency-dependent screened Coulomb interaction is crucial to capture these states; the frequency-dependence of the TDDFT xc kernel extracted from this has two origins, one from the folding of the space variables, and the other from this explicit dependence. The approach however led to spurious excitations, thought to be due to a self-screening error, and later shown to be avoidable  by imposing a condition for number-conservation in the Bethe-Salpeter approach~\cite{SROM11}. The relation between this approach and the propagator approach was clarified in Ref.~\cite{CH16}.

The dressed kernel has been tested on a range of different molecules, computing excited state geometries as well as energies~\cite{MW09,MMWA11,HIRC11}. An extensive study using the development version of the deMon2k code~\cite{deMon2k} on 28 organic molecules suggested that dressed TDDFT gives the best results when the adiabatic kernel it is paired with is a hybrid. 

Still, there are double-excitations for which the dressed kernel does not apply: when the condition that the subspace containing the double-excitation is uncoupled from the others, in particular the ground-state, does not hold. 
When the KS lowest unoccupied molecular orbital energy (LUMO) lies low close to a doubly-occupied highest occupied orbital energy (HOMO), {\it any} single-excitation out of the HOMO will be near-degenerate with a double-excitation where the other electron occupying the HOMO hops into the low-lying LUMO~\cite{MT06} (see also panel c in  (\textbf{Figure \ref{fig:doublesketch}}). This means that $f\Hxc(\omega)$ has a strong frequency-dependence throughout the spectrum. It also means that the dressed kernel is not appropriate since the SPA under which it is to be applied breaks down~\cite{GPG00,GM12chap,M05c}. 
Two relevant situations in which this occurs are conical intersections~\cite{TTRF08} and stretched single-bonds such as in dissociating diatomic molecules~\cite{GGGB00,M05c,MT06} (see Sec.~\ref{sec:CTopen}).  In these cases, ground-state DFT also struggles tremendously because the single Slater determinant character of the KS state is so far from the exact ground-state which is strongly correlated.

\subsubsection{Oscillator strengths}
While the frequencies give the position of the peaks in the absorption spectrum, an aspect of the spectrum that tends to be less discussed is the height of the peaks in the spectrum, i.e. the oscillator strength. TDDFT gives in principle the exact oscillator strengths, extracted from the residues of the susceptibility $\chi$, or the eigenvectors of the matrix $R$ of Eq.~\ref{eq:matrix}~\cite{C95,C96,GPG00,GM12chap}; a recent benchmarking for general excitations in small compounds can be found in Ref.~\cite{SBMLJ20}. Ref.~\cite{C95} showed that frequency-dependence of the xc kernel imposes a renormalization of the eigenvectors $F_j$ of $R$ compared to those obtained from the adiabatic approximation  when computing the oscillator strengths
(Eqs.~4.39 -- 4.41 of Ref.~\cite{C95}). To my knowledge, the effect this renormalization has on shattering the single KS peak strength into the mixed single and double character components has not been explored very much; I know only of one work on an asymmetric Hubbard dimer~\cite{CFMB18}. 

\subsection{Searching for Doubles Elsewhere Within DFT}
\label{sec:delsewhere}
Although dressed TDDFT has successfully computed excitation energies of double excitations, it is not widely used; this is perhaps because it is not applied in a black-box way since one first scans the single excitations out of an adiabatic approximation to see where to apply the frequency-dependent part of the kernel. 
We briefly mention here some other density-functional based approaches that have been explored for double-excitations. 

A natural approach is to consider quadratic response: given that two photons are required to excite two electrons in a non-interacting system, one might hope that an adiabatic approximation used within quadratic response theory has the right structure to couple single and double KS excitations. Unfortunately, it was found~\cite{EGCM11,TC03} that while adiabatic quadratic response does contain poles at the sum of linear-response-corrected KS single excitations, it misses the {\it mixing} between single KS excitations with these double excitations.  Using the Tamm-Dancoff approximation in the linear-response part of the calculation makes even these poles disappear. 

Back to linear response, instead of improving the xc kernel, spin-flip TDDFT instead modifies the reference state around which the linear response is performed~\cite{SHK03, RVA10}; this was originally introduced to access ground states of multi-reference character. A double-excitation with respect to the ground-state appears as a single-excitation of the new reference state. Choosing a high-spin triplet state as reference and applying spin-flip excitations, double KS excitations contribute to the TDDFT linear response using the usual adiabatic xc kernels, or noncollinear ones designed from considering the nature of the reference state~\cite{WZ05}. 

An alternative linear response theory was recently developed,  in a similar spirit to TDDFT, but distinct in that the excitations involve electron additions or removals: In particle-particle random phase approximation (pp-RPA)~\cite{YAY13,YPLY14}, the reference state is the ground-state of $(N-2)$ electrons instead of $N$. Then two electrons are added to any of the unoccupied orbitals of the reference; in this way, double-excitations naturally arise. There are some limitations, depending on the character of the double-excitation (e.g. it does not capture excitations in which a significant amount of the hole is in the HOMO-1), but works well otherwise; also for other difficult excitations in TDDFT including charge transfer.

Falling back to DFT, we note that constrained variational methods have been formulated to reproduce excited states of a given character, and have been applied to double-excitations; in particular $\Delta$SCF and orbital-optimized DFT~\cite{HH21}, constricted variational DFT~\cite{SKZ14},  and eXcited Constrained DFT~\cite{RP18}.

Stepping outside standard KS DFT, a promising approach that has resurfaced in recent years is ensemble-DFT~\cite{GOK88,GOK88b,OGK88}.  Ensemble-DFT is based on the rigorous Gross-Oliveira-Kohn variational principle for ground and excited states, and its initial exposition that pre-dated the linear-response framework of TDDFT.   
Although early approximations for the functionals were not accurate enough to be useful~\cite{GPG02,TN03}, very recent algorithmic and functional developments have rewoken the exploration of whether it could become a practical and accurate method, with computational cost similar to KS DFT. Several works have considered double excitations in this framework~\cite{SB18,LF20,MSFL20,PYTBNU14}.



\section{Charge-Transfer Excitations} 
\label{sec:CT}
A charge-transfer excitation is one in which a large fraction of the excited state electron density is localized in a region with little spatial overlap with the density of the ground-state. This occurs in a number of situations, for example, when the excited state is rotated with respect to the ground-state as in twisted intramolecular charge transfer compounds, or at stretched geometries as in a dissociating bond. In the limit of minimal overlap between the donor and acceptor states, one can obtain the lowest charge-transfer energy staying within a ground-state DFT, by using constrained DFT~\cite{KKV12}. One can extract useful coupling matrix elements from the constrained states. For a general approach to charge-transfer one needs to consider TDDFT.
Charge transfer  plays a key role in many central processes in science, including photosynthesis, photovoltaic devices, molecular switches, nanoscale conductance, reactions in solvents and at interfaces. In many of these applications, the systems are large enough that TDDFT is the only practical option, and so charge-transfer has received a lot of attention.  A detailed review of the issues and developments in both  linear response  as well as for fully time-resolved, non-perturbative dynamics can be found in Ref.~\cite{M17}.

It was realized in the early aughts that standard approximate TDDFT functionals severely underestimate charge-transfer excitation energies~\cite{T03,DH04}. Yet at around the same time, a TDDFT study  made a breakthrough in the explanation of the charge-transfer process responsible for the dual fluorescence of 4-dimethyl-aminobenzonitrile (DMABN) in polar solvents whose mechanism had until then remained a mystery~\cite{RF04}. The nature of the red-shifted emission band had been thought to be due to an intramolecular charge-transfer state, but the lack of accurate but computationally-efficient methods at the time made it difficult to know whether it had a twisted or planar quinoidal structure.  Using TDDFT with the B3LYP functional,  Ref.~\cite{RF04} could without doubt identify the electronic and geometric nature of the state, and the mechanism that led to the dual fluorescence. This was an early success story for TDDFT and charge-transfer processes, where  the  underestimation of the charge-transfer excitation energies  themselves was not important. They were indeed likely underestimated but two aspects of the study meant that it did not affect the conclusions: first, the calculations were performed in the gas phase while the phenomenon occurs in solvent that would tend to lower the excitation energy anyway, and second, the excited properties such as vibrational frequencies and force constants used to identify the state appear to be generally less sensitive. But in general, the large underestimation of charge-transfer excitation energies hampers the predictivity of standard  TDDFT approximations in a range of applications in physics, chemistry, and biology and has driven tremendous developments in the past decade or so, such that now first-principles non-empirical functional approximations are available that can in many cases yield reliable and predictive results for charge-transfer excitations~\cite{K17b}. 

For the ensuing discussion on why TDDFT finds these excitations so challenging, we  first recall what their exact value should be. Consider an excitation of a stretched neutral molecule where one electron has transferred from one end (the donor) to the other (the acceptor). Then at large separations, the exact frequency of this excitation approaches 
\ben
\Omega = I^{D} - A^{A} - 1/R
\label{eq:CTex}
\een
where $I^D = E^D(N_D - 1) - E^D(N_D)$ is the ionization energy of the $N_D$-electron donor, $A^A = E^A(N_A) - E^A(N_A+1)$ is the electron affinity of the $N_A$-electron acceptor, and $-1/R$ is the electrostatic attraction between the fragments after the transfer, lowest-order in the separation $R$.

How these excitations are represented in TDDFT depends on the character of the underlying KS orbitals. We must distinguish two cases: charge-transfer between closed-shell fragments, and charge-transfer between open-shell fragments. The latter case is particularly challenging for TDDFT because of the strongly-correlated nature of the ground-state, and the analysis of the situation is quite distinct from the former. 
We note that several diagnostics of the degree of charge-transfer in an excitation have been useful~\cite{PBHT08,GCMA13,LU16}. 

\subsection{Charge-transfer excitations between closed-shell fragments}
In this case, we have a pair of electrons in the HOMO of the donor from which we transfer one to the LUMO of the acceptor. The Kohn-Sham orbital energy difference is then simply
 \ben
 \omega_q = \epsilon_L^A - \epsilon_H^D\,,
 \label{eq:KSCTex}
 \een
 and the TDDFT procedure of Eqs.~\ref{eq:matrix} provides a diagonal correction and mixes this excitation with other excitations through the $f\xc$-matrix element that goes into $R_{qq'}$. However,  it is evident from this equation that the KS transition-density $\Phi_q(\br) = \phi_H^{D*}(\br)\phi_L^A(\br)$ is exponentially small as a function of $R$, so for a non-vanishing correction to the KS orbital energy difference, $f\xc(\br, \br',\omega)$ must exponentially grow as a function of $R$. Local and semi-local functionals (LDA/GGA) do not have this property, so their TDDFT excitation energy collapses to the orbital energy difference. 

One might wonder, how far is the KS orbital energy difference $\omega_q = \epsilon_L^A - \epsilon_H^D$ from the exact CT excitation energy Eq.~\ref{eq:CTex}? 
The answer depends not only on what  ground-state functional is being used, but also on whether the calculation is performed within pure KS DFT or the generalized KS framework~\cite{GNGK20,SGVML96,GL97}. There is a key difference between these two formalisms that has a significant consequence for charge-transfer excitations: in the former, unoccupied orbital energies are excitations of the neutral system, while in the latter, they have a character somewhere in between neutral and addition energies depending on the amount of Hartree-Fock that is mixed in.  Given that charge-transfer excitations, albeit overall neutral, do involve the addition of one electron on one moiety, the generalized KS may have a practical advantage. In the following, we will briefly outline different approaches, considering the nature of the bare KS orbital energy differences as well as the TDDFT correction from $f\xc$, which, for non-local functionals could be non-vanishing.
We will consider several distinct and contrasting approaches, and again note that more detailed exposition is given in Ref.~\cite{M17}. 

First, if the {\it exact} ground-state functional was somehow known and used, then we have
$\omega_q = I^D - A\s^A = I^D - A^A - \Delta\xc^A$, since in DFT the magnitude of the HOMO orbital energy is exactly equal to the true ionization energy but the KS LUMO orbital energy differs from the electron affinity by the derivative-discontinuity $\Delta\xc^A$
~\cite{GKKG00,PL83, SS83, P85b, AB85, PPLB82,M17}. That is, the exact KS orbital energy difference is lacking relaxation contributions to the acceptor's electron affinity as well as the $-1/R$ behavior at large separations; 
if the exact ground-state functional was used in the TDDFT calculation, these terms would need to result from the $f\xc$ term in Eq.~(\ref{eq:matrix}), and, from the above discussion, we see this requires $f\xc(\br,\br',\omega)$ to have some matrix elements that grow exponentially with fragment separation $R$ (note, this is {\it not} the same as growing with $\br - \br'$). 

Of course the exact ground-state functional is not known, so the second case we consider is the situation for local and semi-local functionals, {\it LDA/GGA}\begin{marginnote}[]\entry{LDA; GGA}{local density approximation; generalized gradient approximation}\end{marginnote}. With local approximations, the $\epsilon_H$ in Eq.~\ref{eq:KSCTex}  is a significant underestimate of the ionization energy: because the LDA/GGA potentials depend only (semi)locally on the density, and the density falls exponentially with the distance from the atom,  the LDA/GGA potentials go to zero exponentially instead of having the slower $-1/r$ tail away from a finite system. Although this does not affect the  lower energy orbitals occupied in the ground-state so severely,  the valence levels that probe these regions further from the atom, get pushed upwards and hence the LDA/GGA HOMO (and LUMO) orbital energies are too small.  Tozer showed that the error for charge-transfer excitations when using these functionals tends to the average of the derivative-discontinuities of the donor and acceptor~\cite{T03}:
$\omega_q = I^D - A^A -1/2(\Delta\xc^D + \Delta\xc^A)$. 
This is unchanged by the TDDFT correction from $f\xc$ due to the local nature of the kernel in  LDA/GGA. As a fix, configuration-interaction singles (CIS) was added to simply shift the LDA/GGA values in Ref.~\cite{DWH03}; CIS alone gives the $-1/R$ behavior but tends to the Hartree-Fock orbital energy difference which gives an overestimate. A fix that stays within TDDFT was provided in Ref.~\cite{GB04} applying a kernel that switches on an asymptotic correction to ALDA when the $f\xc$ matrix element becomes too small.

The third case is, in a sense, a functional approximation at the opposite extreme of DFT: exact-exchange ({\it EXX})\begin{marginnote}[]\entry{EXX:}{exact exchange ({\it not} Hartree-Fock)}\end{marginnote}. In contrast to LDA, this has a non-local dependence on the density (not to be confused with still providing a local multiplicative potential).  EXX has a fundamental importance complementary to that of LDA, in that it results from first-order  (G\"orling-Levy) perturbation theory in the electron-interaction with respect to the KS system~\cite{GL93,GS99}. 
Since the EXX potential does have the correct $-1/r$ behavior far from a finite system, the KS HOMO orbital energy approximates the true ionization energy much better than LDA does. Further,  through orbital-dependence, the TDEXX kernel contains the required diverging property as a function of $R$~\cite{HIG09, GIHG09,HG12,HG13}, yielding both the exchange-component to the derivative-discontinuity $\Delta\x$ as well as the $-1/R$. Frequency-dependence is an important aspect: if the adiabatic EXX kernel was used, $f\xc(\omega = 0)$, instead the correction vanishes as $R \to 0$~\cite{HG12}. The EXX kernel must be evaluated at the charge-transfer excitation energy in order to yield a finite correction; this is related to the strong frequency-dependence of the derivative-discontinuity of the xc kernel~\cite{MK05,HG12}. 

We turn next to {\it global hybrid} functionals\begin{marginnote}[]\entry{global hybrid:}{fraction of Hartree-Fock exchange $+$ a (semi)local functional}\end{marginnote}
, popular throughout quantum chemistry. These functionals combine a fraction of Hartree-Fock exchange with a local or semi-local functional; for example, with the ubiquitous B3LYP, the fraction is $a\x = 1/4$. Hybrid functionals fall within the generalized KS formulation~\cite{GNGK20,SGVML96,GL97}, where the formal justification arises from including a fraction of the electron-electron interaction $W$ in the minimization of $T + a\x W$ over Slater determinants that yield a fixed density; the pure KS DFT approach on the other hand minimizes purely the kinetic energy $T$. The generalized KS potential is no longer identical for each orbital, and is non-local (i.e. non-multiplicative), as it includes a fraction of the Hartree-Fock potential. As a result, the HOMO orbital experiences a $-a\x/r$ potential asymptotically away from a finite system instead of the exponential fall-off of (semi)local functionals, so its energy is not as badly underestimated. Further, the LUMO eigenvalue includes this fraction of the exchange contribution to the derivative-discontinuity~\cite{SGVML96}; this reflects the partial affinity nature of the unoccupied levels in a hybrid, since in Hartree-Fock the unoccupied orbitals ``see" an $(N+1)$-electron system, while in pure KS DFT, they see an $N$-electron system. This also underlies the difference between EXX performed with an optimized effective potential and Hartree-Fock.
Thus, hybrids reduce the underestimation of the orbital energy difference for a charge-transfer excitation. Additionally, the Fock-exchange contributes as  $-a\x/R$ at large separations to the $f\xc$ correction.
That is, the KS orbital energy difference provides a partial derivative-discontinuity, while the $f\xc$ correction partially provides the asymptotic behavior with $R$.

The global hybrid nudges us towards the correct excitation energy but the {\it range-separated hybrid (RSH)}\begin{marginnote}[]\entry{RSH}{range-separated hybrid includes full Hartree-Fock exchange at large range}\end{marginnote}
 takes us further by recovering the full Hartree-Fock exchange at large electron-electron separation~\cite{LSWS97,SS85}. The idea is to split the Coulomb interaction into a long-range and short-range term, such as
\ben
\frac{1}{\vert\br_1 - \br_2\vert} =\frac{ \rm{erf}(\gamma\vert\br_1 - \br_2\vert)}{\vert\br_1 - \br_2\vert} + \frac{ 1-{\rm erf}(\gamma\vert\br_1 - \br_2\vert)}{\vert\br_1 - \br_2\vert}
\een
and use local or semi-local approximation for the second term which dominates at short distances and dies off at long range, while using Hartree-Fock for the first term which dominates at long range and dies at short-range. The range-separation parameter $\gamma$ controls the distance at which the long-range part begins to take over: the larger the $\gamma$ is, the smaller the distance at which the Hartree-Fock kicks in. This approach balances the advantages of the Hartree-Fock and semilocal DFT worlds, capturing dynamical correlation and taking advantage of the error cancellation between exchange and correlation from semi-local DFT at short-range, while using Hartree-Fock for the long-range interaction that is dominated by exchange and poorly captured by semilocal DFT. Variations of this essential idea include using also some Hartree-Fock at short-range; the CAM-B3LYP combines RSH with B3LYP including a third parameter in the range-separation such that there is a non-uniform fraction of Hartree-Fock exchange at all separations~\cite{ITYH01,YTH04,HJS08}.
For the problem of charge-transfer, RSH yields the exact $-1/R$ dependence at large $R$, and gives an approximate discontinuity correction to the LUMO orbital energy, moving it towards the physical electron affinity of the donor~\cite{TTYY04}. A challenge is in the empiricism: finding parameters that yield a balanced description of charge-transfer as well as local valence and Rydberg excitations as well as ground-state properties. The results can be very sensitive to these parameters~\cite{K17b}. Also different range-separated forms have been explored, including with density-dependent parameters~\cite{BN05,BLS10}; some forms tend to yield a more uniform performance than others~\cite{RMH09, HJS08b}.  

To avoid empiricism completely, Baer, Kronik, and co-workers developed {\it optimally-tuned RSH}, where the range-separation parameter is chosen to minimize the difference between the ionization potential of the donor and the donor's HOMO eigenvalue as well as the difference between the electron affinity of the acceptor and the acceptor's LUMO eigenvalue, all determined consistently with the same functional~\cite{KSRB12,BLS10,SKB09,KB14}. The 
method is arguably the most predictive of the different approximations for charge-transfer excitations between closed-shell fragments, and also captures local outer-valence excitations well~\cite{EWRS14}.
Still, there are several issues with RSH that should be borne in mind. The potential energy surfaces for triplets and singlets can show erratic zig-zagging behavior, due to the tuning;  this does not happen when the range-separation parameter is fixed~\cite{KKK13}.
RSH violates size-consistency~\cite{KKK13,KB14}, and
further,  the values for the RSH parameter found by optimal tuning to the ionization potential and electron affinity, tend not to give good ground-state binding~\cite{SKMKK14}.

Finally, other approximations that have shown some degree of success in capturing charge-transfer excitations, include the self-interaction corrected LDA~\cite{HKK12,HK12}, applied within a generalized time-dependent optimized effective potential framework~\cite{KKM08}. As pointed out in Ref.~\cite{P90}, the appearance of a finite derivative-discontinuity  in a functional is related to its correction of self-interaction. The heavy computational cost of this approach has limited its application. On the other hand, a  less expensive approach that was recently shown to have promise with charge-transfer excitations is the TASK meta-GGA~\cite{AK19}: non-locality arises through its orbital dependence such that  it yields response properties similar to exact exchange but without the numerical cost, and gives some improvement of medium-range charge-transfer excitation energies~\cite{HK20}.
Double-hybrid functionals that combine a second-order correlation part to GGA for correlation on top of a usual hybrid functional have been explored~\cite{GN07}, and so have highly-parameterized functionals such as M06-HF meta-GGA and MN15 that use many parameters fit to datasets~\cite{ZT06,YHLT16}.

\subsection{Charge-transfer excitations between open-shell fragments}
\label{sec:CTopen}
The analysis of the charge-transfer problem for TDDFT is quite distinct from the previous case when the neutral molecule is composed of open-shell fragments, such as in a heteroatomic diatomic molecule. There is a fundamental difference in the nature of the KS HOMO and LUMO compared to the closed-shell fragments case that makes the previous analysis  not applicable. With open-shell fragments, the exact singlet ground-state  has a doubly-occupied HOMO orbital that is delocalized over both fragments, quite in contrast to the localized HOMO of the closed-shell case. The LUMO is also delocalized, and its orbital energy becomes degenerate with the HOMO in the limit of infinite separation. The static correlation in the KS system means that at large separations, the ground-state KS Slater determinant has a fundamentally different structure to the interacting wavefunction which has a Heitler-London form, in contrast to the closed-shell fragment case.

While the ground-state situation is pathological for approximate TDDFT, it is an important one for systems away from equilibrium, such as in bond-breaking, and the excitations are essential to get right for accurate dynamics in photo-dissociation processes, for example. 

The exact ground-state KS potential of the widely-separated molecule is locally similar to that of the atoms in the vicinity of each atom but has a step in between of a size approaching the difference in the ionization potentials of the two atoms (see panel c of \textbf{Figure \ref{fig:doublesketch}} and Sidebar~\ref{sidebar:step})~\cite{P85b,TMM09,AB85,GB96,HTR09}. This makes the atomic HOMO's ``line up" in order for the molecular HOMO to correctly straddle both atoms and capture the correct ground-state density. However the interatomic step is not captured by approximate functionals: the semilocal molecular HOMO is delocalized over both atoms but does not reduce to the combination of the atomic HOMOs, and the molecule dissociates unphysically to fractionally charged species. 
Hybrid functionals, including RSH, also do not dissociate correctly into the neutral atomic species. This is true for general $N$ and can be seen simply by considering a model heteroatomic two-electron system where there is one electron on each atom: for two electrons, the Hartree-Fock exchange potential is $ -v\H/2$, so that in the vicinity of each atom in the widely-separated limit, $v\s = v\ext + v\H/2$ while it should be $v\ext$. The local atomic densities are wrong. 
A functional that was inspired by density-matrix functional theory with explicit dependence on both occupied and virtual orbitals has been shown to capture the step structure~\cite{BB02,GB06b}, but whether this can be turned into a practical approach remains to be seen. 
Static correlation is well-known to be a difficult regime for density functional approximations in the ground state, and its implication for excitation energies and response are very challenging; some recent progress based on the strictly-correlated electron approach for dissociation can be found in Refs.~\cite{GVG18,JSGG19,GG20}.

All KS excitations of such a system are near-degenerate with a double-excitation where both electrons are excited out of the HOMO: an additional excitation from the HOMO to the LUMO adds very little cost to a single-excitation from the HOMO to any unoccupied orbital. Both HOMO and LUMO are delocalized over both atoms and have substantial overlap, with  an orbital energy difference that vanishes exponentially with the separation $R$~\cite{M05c,MT06,M17}; but these are the orbitals involved in the lowest charge-transfer excitations of the molecule. 
Since the bare KS excitation energy $\omega_q$ between the HOMO and LUMO vanishes exponentially with separation $R$, the TDDFT corrections in the matrix elements involving $f\xc$ are responsible for the entire charge-transfer energy of Eq.~\ref{eq:CTex}. Further, if we were to simplify the analysis through considering only the HOMO-LUMO excitation subspace, we observe
Eq.~\ref{eq:SPA} is not valid, since the $\omega_q$ is smaller than the correction. Within the SMA (Eq.~\ref{eq:SMA}), the exact $f\xc$ matrix element has a very strong frequency-dependence and diverges exponentially with the fragment separation $R$~\cite{M05c,MT06,M17}.  The exact-exchange kernel~\cite{HG12,HG13} discussed briefly in the previous section displays a frequency-dependent divergence with respect to $R$, but in the present case, the divergence occurs in the correlation potential. In fact, throughout all frequencies, the xc kernel in the present case is rife with strong-frequency dependence. This can be understood as 
 due to mixing with  the near-degenerate double-excitations throughout the spectrum. Physically this mixing is essential to avoid yielding excited states that have ``half" an electron excess or deficient on one atom (see also the figure).

\begin{textbox}[h]
\section{Step in the ground-state KS potential}
\label{sidebar:step}
An explicit demonstration of the step in the ground-state KS potential for the case of open-shell fragments follows from simply considering the KS equation for the external potential
\ben
v\ext(\br) = v_a(\br) + v_b(\br)
\een
where $v_{a}(\br) (v_{b}(\br)) $ represents the electron-nuclear potential for atom $a (b)$ localized around $r = -R/2 (R/2)$.
The potentials $v_a(\br)$ and $v_b(\br)$ each support an odd number of electrons in their atomic ground-states, and, for large separations $R$, the HOMO orbital of the diatomic molecule has the form 
\ben
\phi^H(\br) = \sqrt{\left(\vert\phi_a^H(\br)\vert^2 + \vert\phi_b^H(\br)\vert^2\right)/2}
\een
with $\phi_a^H (\phi_b^H)$ the atomic HOMO of atom $a (b)$ localized at $-R/2 (R/2)$.

Consider now the ground-state KS equation in the vicinity of atom $a$ when the intermolecular separation $R$ is large: here $v\ext(\br)\approx v_a(\br) -Z_b/2R, v\H[n](\br) \approx v\H[n_a](\br) + N_b/2R$ where $Z_b$ and $N_b$ are atom $b$'s nuclear charge and electron number. Then, for neutral atoms, the KS equation for the molecular HOMO is
\ben
\mbox{\rm near atom} \, a, \; \left(-\nabla^2/2 + v_a(\br) + v_H[n_a](\br) + v\xc[n](\br)\right)\phi^H(\br) = \eps^H\phi^H(\br), \;\;\; \phi^H(\br) \approx \phi^H_a(\br)/\sqrt{2}
\een
But we also know that the atomic HOMO satisfies the atomic KS equation
\ben
\left(-\nabla^2/2 + v_a(\br) + v_H[n_a](\br) + v\xc[n_a](\br)\right)\phi^H_a(\br) = \eps_a^H\phi^H_a(\br)
\een
which means that the difference in the xc potential of the molecule compared with that of the atom  is
\ben
\mbox{\rm near atom} \, a, \; \Delta v\xc[n](\br \approx -R/2) \equiv v\xc[n](\br) - v\xc[n_a](\br) = \epsilon^H - \eps^H_a
\een
The same argument applied near atom $b$ leads to
$
\Delta v\xc[n](\br \approx R/2) \equiv v\xc[n](\br) - v\xc[n_b](\br) = \epsilon^H - \eps^H_b
$. 
Thus, across the molecule, there is a step in the difference of the molecular xc potential compared to the atomic xc potentials:
\ben
\Delta v\xc[n](\br \approx R/2) - \Delta v\xc[n](\br \approx -R/2)  = \epsilon_a^H - \epsilon_b^H = I_b - I_a
\een

An illustration of the ground-state KS potential and orbitals is shown in panel c of \textbf{Figure \ref{fig:doublesketch}}.


\end{textbox}

\section{Outlook} 
\label{sec:outlook}
Although at equilibrium geometries, double excitations may be relevant in relatively few situations, they are absolutely crucial in coupled electron-ion dynamics following an excitation or driven by a laser. It was pointed out by Levine and co-workers~\cite{LKQM06} that even if problematic excitations are not present at the equilibrium geometry, that  photo-induced dynamics traverses large ranges of nuclear configurations and the likelihood of curve-crossing means that challenging excitations such as double and charge-transfer excitations, and conical intersections are likely to be encountered. For example, in ethylene, a $\pi\to \pi^*$ excitation is followed by a twisting and pyramidalization. The global minimum on $S_1$ is of doubly-excited character and at a geometry that is both twisted and pyramidalized, but the absence of double-excitations in adiabatic TDDFT yields an $S_1$ minimum that is purely twisted~\cite{LKQM06}. This would clearly alter the predictions of the coupled electron-nuclear dynamics. 

Likewise, long-range charge-transfer excitations of a molecule are highly relevant for the dynamics of a molecule following a photo-excitation out of the ground-state of its equilibrium geometry, or driven by a laser field. In these situations,  the propensity for dissociating is increased, for either the closed-shell or  open-shell fragment case. The coupled electronic and nuclear motion straddles several Born-Oppenheimer potential energy surfaces, and for TDDFT to be  used reliably in mixed quantum-classical Ehrenfest or surface-hopping calculations, the surfaces obtained from linear response TDDFT must be globally accurate in order to get accurate dynamics. Indeed, the lack of computationally efficient and reliably accurate electronic structure methods is arguably the main hindrance to non-adiabatic dynamics calculations, more so than the choice of method used for the electron-nuclear correlation: all the methods used to couple the electronic and nuclear motion 
would greatly benefit from improved functional approximations in TDDFT, especially for for double-excitations, charge-transfer excitations, and conical intersections. Whether developments uncovering and modeling the memory-dependence of exact functionals in the real-time domain can lead to new approximations for linear response remain to be seen~\cite{LM20b}. 
Out of wavefunction methods that may be more reliable for these excitations, one would wish to avoid issues such as inadequacy of a chosen active spaces as the molecule explores geometries where the electronic structure has significantly changed character or is not well understood, e.g. in CASSCF and related methods, which, at the same time, are limited to much smaller molecules than possible with TDDFT; even more so with CC methods. Even if possible with expanding computational architectures, running expensive calculations that use a lot of computational power should be carefully justified  in the current climate-crisis. The problem also urges forward the further exploration of other reduced-variable theories, including the ensemble-DFT (discussed in application to double-excitations in Sec.~\ref{sec:delsewhere} has also been applied for charge-transfer excitations~\cite{GKP18,FF14}), Green's function and Bethe-Salpeter methods~\cite{JDB17,LKYRAKB20,Sahar18,LB20}, and one-body density-matrix functional theory~\cite{GBG08,GPGB09,GGB10}.
Future years hope to see further significant progress continuing from that made in recent decades, with ever-improving numerical implementations, and new and exciting applications yet to be imagined.

\begin{summary}[SUMMARY POINTS]
\begin{enumerate}
\item  TDDFT provides an elegant and rigorous way to obtain electronic excitations and response for many-electron systems, that has achieved an unrivaled balance between accuracy and efficiency. The acrobatics of the functionals involved that allow non-interacting electrons to reproduce the exact density of an interacting system remains an intriguing, important, and fun research area.
\item  There are certain excitations for which the standard functionals that are semi-local in space and local in time in their dependence on the density do not perform well, such as double-excitations and charge-transfer excitations.
\item There has been significant process in developing functionals with improved accuracy and reliability over the recent years that address these challenging excitations in a non-empirical way. For double-excitations, a strong frequency-dependence is required in the xc kernel, while for many, but not all, classes of charge-transfer excitations, spatial non-locality is more important. 
\end{enumerate}
\end{summary}

\begin{issues}[FUTURE ISSUES]
\begin{enumerate}
\item  We can look forward to further progress in turning the recent developments into black-box methods, from the algorithmic and computational point of view (e.g. practical treatment of frequency-dependent functionals) and in developments to ease the computational efficiency to benefit photochemical dynamics applications with the more accurate functionals. 
\item  Relatedly, the treatment of conical intersections with the ground-state remains a challenge. 
\item  Understanding the performance of the full response properties beyond merely the value of the excitation energies (e.g. oscillator strengths), from the newer functionals will lead to more food for thought.
\item  The exactness of the underlying theory offers hope that further development of first-principles functionals from different starting points may lead to improved future robust, reliable and predictive functional approximations. 
\end{enumerate}
\end{issues}

\section*{DISCLOSURE STATEMENT}
The author is not aware of any affiliations, memberships, funding, or financial holdings that
might be perceived as affecting the objectivity of this review. 

\section*{ACKNOWLEDGMENTS}
This review is dedicated to Professor Bob Cave, one of the kindest and most brilliant scientists ever, whom we continue to deeply miss. 
Financial support from the National Science Foundation Award CHE-1940333 and from the Department
of Energy, Office of Basic Energy Sciences, Division of Chemical
Sciences, Geosciences and Biosciences under Award No. DESC0020044 is gratefully acknowledged.
%


\bibliographystyle{ar-style3.bst}

\bibliography{./ref.bib}

\end{document}